# Numerical Study on Flow and Heat Transfer of Water and Liquid Metals Within Micro-Scale Heat Sinks for High Heat Dissipation Rate Applications

**Mahyar Pourghasemi** [1]
University of New Mexico
Albuquerque, New Mexico, USA
mpourghasemi@unm.edu

**Nima Fathi** [2]
University of New Mexico
Albuquerque, New Mexico, USA
nfathi@unm.edu

**Sal Rodriguez** [3]
Sandia National Laboratories
Albuquerque, New Mexico, USA
sbrodri@sandia.gov

## ABSTRACT

Compact and small-scale heat exchangers can handle high heat dissipation rates due to their large surface area to volume ratios. Applications involving high heat dissipation rates include, but are not limited to, compact microelectronic processing units, high power laser arrays, fuel cells, as well as fission batteries. Low maintenance cost, small size and dimensions, as well as high convective heat transfer coefficients, make micro-scale heat sinks an efficient and reliable cooling solution for applications with high heat dissipation rates. Despite these advantages, the large pressure drop that occurs within micro-scale heat sinks has restricted their utilization. Slip at the walls of microchannels has been reported to reduce friction factor up to 30%, depending on the hydraulic diameter of the microchannel. Numerical investigations are conducted to comprehensively investigate the effect of slip at walls on friction factor and Nusselt number of liquid flows in micro-scale heat sinks. At the same mass flow rate and inlet Reynolds number, obtained results suggest that slip length on the order of 2 μm enhances the overall thermalhydraulic performance of micro heat sinks by almost 6% in comparison with no-slip boundary condition. 4% increase is observed in channel average Nusselt number while pumping power reduces by 8% in comparison with no-slip boundary condition.

---

[1] Graduate Student at UNM
[2] Presenting/Corresponding Author; Research Faculty
[3] Principal Member of the Technical Staff, Advanced Nuclear Concepts, Org. 8841

# 1    INTRODUCTION

Miniaturised and micro-scale heat sinks can handle high heat dissipation rates within a small space due to their large surface area to volume ratios. Although micro-scale heat sinks offer high convective heat transfer coefficients with low maintenance cost, their application has been restricted due to large pumping power required to circulate coolant along these compact heat exchangers. Induced slip at the walls of micro heat sinks has been observed to reduce friction factor up to 30% [1]–[3]. Large number of researches in the literature has been carried out on gas flow and heat transfer in micro-scale devices with slip boundary condition. However, a few studies have been done to investigate the effect of slip boundary condition on heat transfer rate of liquid flows in micro-scale heat sinks. Sohankar et al. [4] investigated flow and heat transfer of water in a U-shaped microchannel with induced slip at the walls. Reported results showed a 35% decrease in pressure loss and up to 43% enhancement in heat transfer rate at a slip length of 10 μm compared to a no-slip boundary condition. Jing et al. [5] conducted numerical simulations to investigate the effect of induced slip on water flow and heat transfer within microchannel heat sinks with an elliptical cross section. Reported results showed that the effect of slip boundary condition became more significant at smaller hydraulic diameters. An apparent slip length of 1 μm in an elliptical micro-scale heat sink with a hydraulic diameter of 30 μm was observed to decrease the pressure loss by almost 20%. Obtained heat transfer rates were slightly higher for a slip boundary condition compared to no-slip, in agreement with previous reported results in the literature [6]. Hajmohammadi et al. [7] investigated water flow and heat transfer in a silicon based micro heat sink with a slip boundary condition at the walls. Results revealed that the minimum thermal resistance is lower when there is a velocity slippage on the microchannel wall. The obtained optimum aspect ratio for the investigated micro-scale heat sink was observed to increase with increased slip length.

Fluids such as water, ethanol and dielectric liquids have been widely used in single phase forced convection cooling applications due to their relatively high specific heat capacities. Moreover, these liquids are compatible with most materials used to fabricate compact and micro scale heat exchangers. However, these liquids have relatively low thermal conductivity in the range of 0.07-0.7 W/mK and low boiling points in the range of 34-100 ºC. High mass flow rates within micro-scale heat sinks are required to prevent the water temperature to reach its boiling point in applications with heat dissipation rates of 100 W/cm$^2$ or higher. This may cause a non-desirable pressure loss on the order of 30 MPa in micro-scale heat sinks [8], [9].

Moreover, it is not feasible to use water, methanol and deictic liquids in applications with high operating temperature range due to their low boiling points. Liquid metals offer very high thermal conductivity in the range of 16-30 W/mK, while their dynamic viscosities are only slightly higher than water. They can be also used in real world applications with an operating temperature range from 25 ºC to as high as 900 ºC due to their very high boiling points. Luo and Liu [10] conducted experiments to investigate heat transfer of GaInSn within compact and miniaturized heat sinks with a hydraulic diameter range of 0.91 to 2.87 mm and length of 40 mm for microelectronics cooling applications. Obtained Nusselt numbers were observed to increase with increased Peclet number in the laminar flow regime.

A new correlation was proposed to estimate Nusselt number as a function of Peclet number. Muhammad et al. [11] performed numerical simulations to study laminar flow and heat transfer for liquid metals GaIn, GaSn, EGaIn, EGaInSn within a miniaturized heat sink with width of 1 mm, height of 4 mm and length of 40 mm. Reported results showed that EGaIn resulted in the lowest pressure loss while EGaInSn required the highest pumping power.

Moreover, GaIn liquid metals showed the best heat transfer performance due its higher convective heat transfer coefficient. 3D numerical simulations are conducted in this work to investigate the effect of slip on friction factor and Nusselt number of liquid flows in micro-scale heat sinks. Conducted numerical analyses include hydraulic diameter range of 250-400 µm and Reynolds number of 300 to 1900. Moreover, thermal performance of a micro-scale heat sink utilizing NaK as the coolant is assessed for applications involving high operating temperature range of 400 K. Obtained Nusselt numbers were presented and discussed as a function of inlet Peclet number.

## 2    NUMERICAL APPROACH

The Fluent CFD solver was used to numerically solve the continuity, the momentum and the energy equations with mass flow inlet and pressure outlet boundary conditions. The governing equations are discretized using a finite volume method on a collocated grid scheme. Discretized equations are solved using a quasi-steady coupled solver scheme while a second-order upwind interpolation scheme is applied to convection terms. At the interface between solid base material and microchannel walls the continuity of heat flux and temperature are imposed using the following boundary conductions, Eq (1).

$$T_f = T_s; \qquad k_f \nabla T_f = k_s \nabla T_s \tag{1}$$

Figure 1 shows the schematic of conjugate heat transfer and boundary conditions within the modelled micro-scale heat sink.

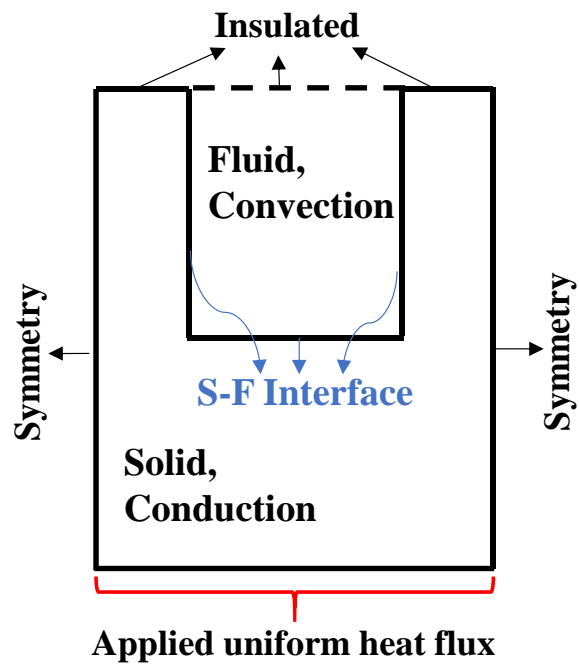

Figure 1: Schematic of applied boundary conditions as well as conjugate heat transfer at interfaces in a microscale heat sink

The following procedure is used to calculate the local and average Nusselt numbers. At any location z, along the microchannel/microtube the average fluid bulk temperature, average wall temperature and average wall heat flux are calculated through Equations (2)-(4).

$$T_b(z) = \frac{\int \rho u C_p T_f dV}{\int \rho u C_p dV} \tag{2}$$

$$T_w(z) = \frac{\{\oint T_w(x,y,z)ds\}_{walls}}{\{\oint ds\}_{walls}} \tag{3}$$

$$q_w(z) = \frac{\{\oint q_w(x,y,z)ds\}_{walls}}{\{\oint ds\}_{walls}} \tag{4}$$

The average local Nusselt number at any location z, along the microchannel is then calculated through Equation (5).

$$Nu(z) = \frac{q_w(z) D_h}{k_b(z) [T_w(z) - T_b(z)]} \tag{5}$$

where, $D_h$ is the microchannel hydraulic diameter and $k_b(z)$ is the fluid thermal conductivity that is evaluated at the local fluid bulk temperature. The average Nusselt number for whole heat sink is calculated by taking the integral of local Nusselt number over the heat sink length of L, using Eq. (6).

$$Nu_{ave} = \frac{\int_0^L Nu(z)dz}{L} \tag{6}$$

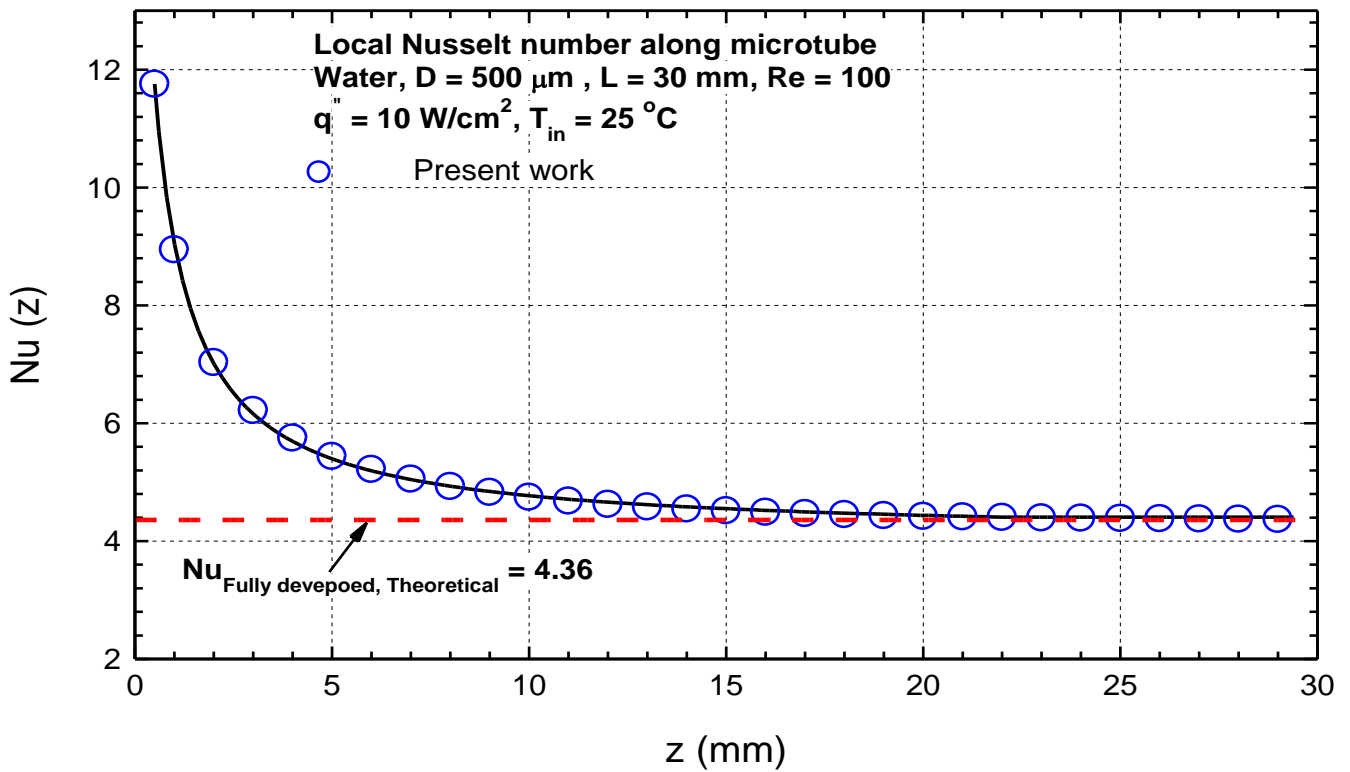

Figure 2: Numerical results for local Nusselt number along a microtube with uniform applied heat flux

Figure 2 illustrates the calculated local average Nusselt number along a microtube using Eqs. (1)-(5). Water is flowing within a microtube of diameter 500 µm and length of 30 mm at applied uniform heat flux of 10 W/cm2. The local Nusselt number is high in the region near the microtube inlet as the liquid flow is still hydrodynamically and thermally developing. The calculated local Nusselt number decreases and approaches to its fully developed theoretical value of 4.36 as water moves toward the microtube outlet. The obtained results confirm the accuracy and reliability of the implemented numerical approach to calculate the local and average Nusselt numbers.

## 2.1 Effect of induced slip on the thermal efficiency of microchannel heat sinks

The geometry of the microchannel heat sink used by Lee et al. [12] in their experiments is modelled in this section to evaluate the effect of induced slip on the thermal efficiency of micro-scale heat sinks. Figure 3 illustrates the schematic of the copper microchannel heat sink. Water is used as the working fluid with an applied heat flux of 40 W/cm². The microchannels have a height of 884 µm, width of 194 µm and length of 25.4 mm. 3D conjugate numerical simulations have been performed with and without slip boundary conditions at Reynolds number range of 500-1000. Figure 4 compares the obtained average Nusselt numbers with experimental data at no slip boundary condition. Excellent agreement between numerical results and experimental data further confirms the accuracy and reliability of the applied numerical methodology to study flow and conjugate heat transfer within micro-scale heat sinks. Previous experimental studies on liquid fluid flow within micro-scale heat sinks utilizing hydrophobic/superhydrophobic surfaces showed pressure drop reduction due to liquid slippage on the walls[2], [13]–[15]. The Maxwell slip model have been widely used in the literature to model fluid slippage at the wall of hydrophobic surfaces [4], [7], [16]. The slip boundary condition in Maxwell model is imposed on the microchannel walls using apparent slip length (β) constant and the velocity gradient normal to the walls as shown in equation 7 [6], [16].

$$u_{slip} = \beta \, \frac{\partial u}{\partial n} \tag{7}$$

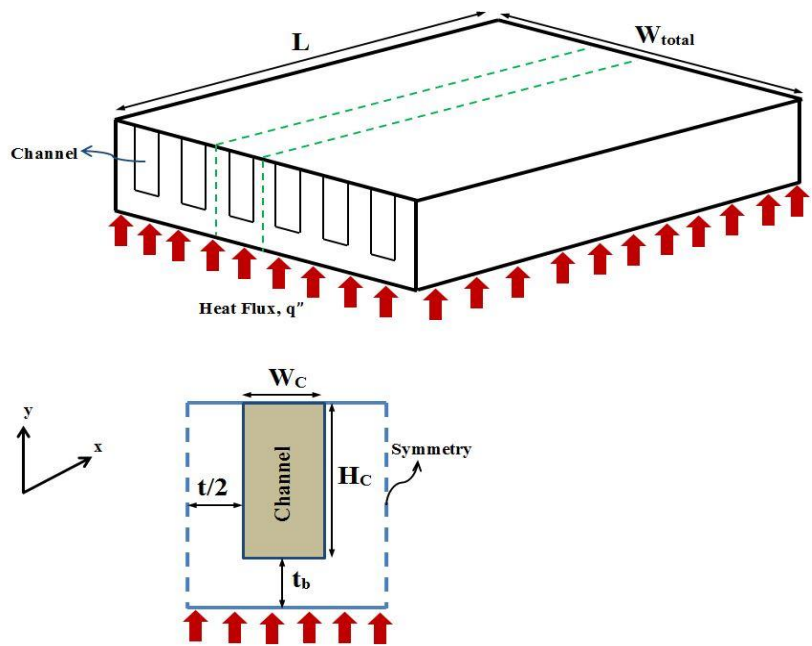

Figure3: Schematic of microchannel heat sink used in present numerical simulations

As can be seen from the obtained results in Figure 5 (a)-(b), induced slip at the microchannel walls tend to reduce the pressure drop while enhancing the heat transfer rate. Water slips at the microchannel walls, enhancing the convective heat transfer rate, which leads to higher Nusselt numbers. For example, at Reynolds number of 1000, the average Nusselt number increases from 9.04 at slip length of $\beta = 0$ (no-slip) to 9.39 at $\beta = 2$ µm. On the other hand, pressure drop along the microchannel decreases from 19.77 to 18.3 kPa, respectively. Overall, a small, induced slip on the order of 2 µm can enhance the hydrodynamic and thermal performance of the investigated microchannel heat sink by almost 8% and 4 %, respectively.

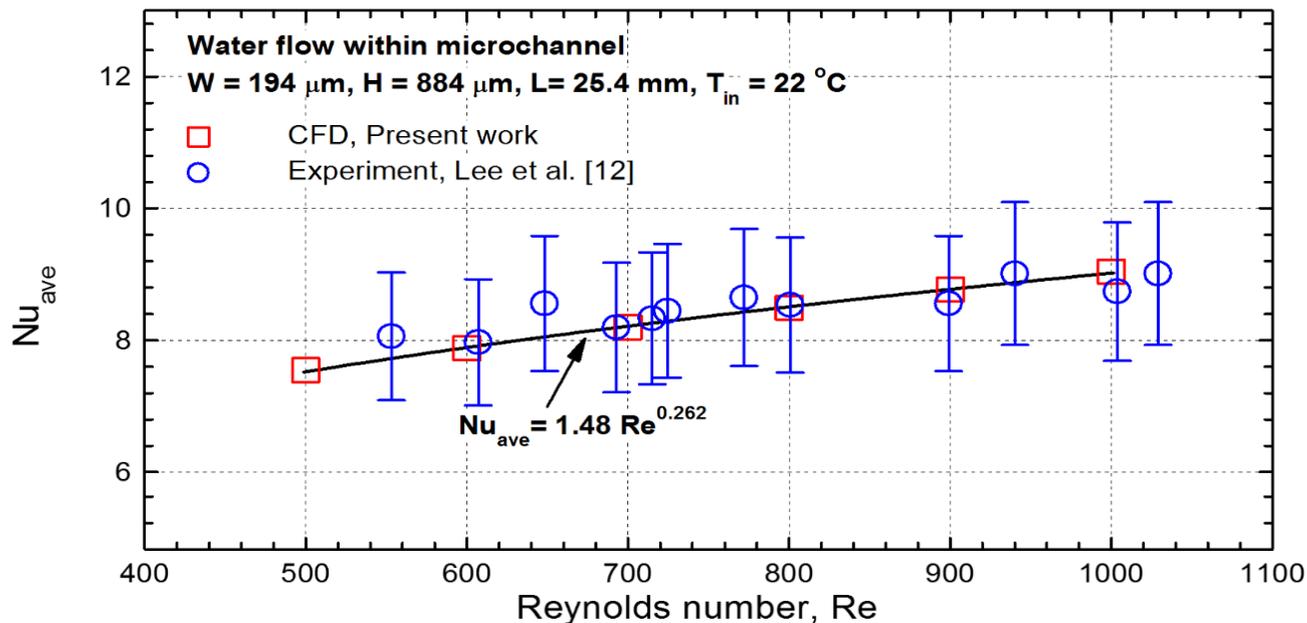

Figure 4: Average Nu from the CFD vs experimental data with no-slip boundary conditio

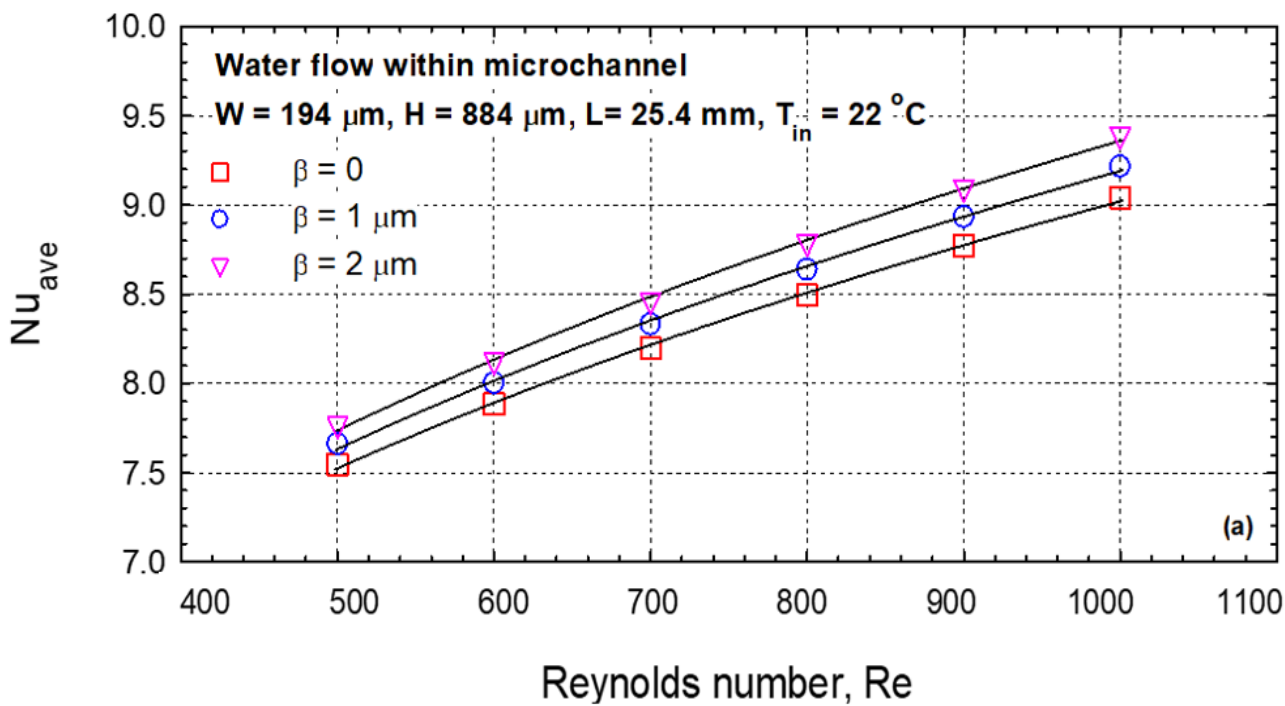

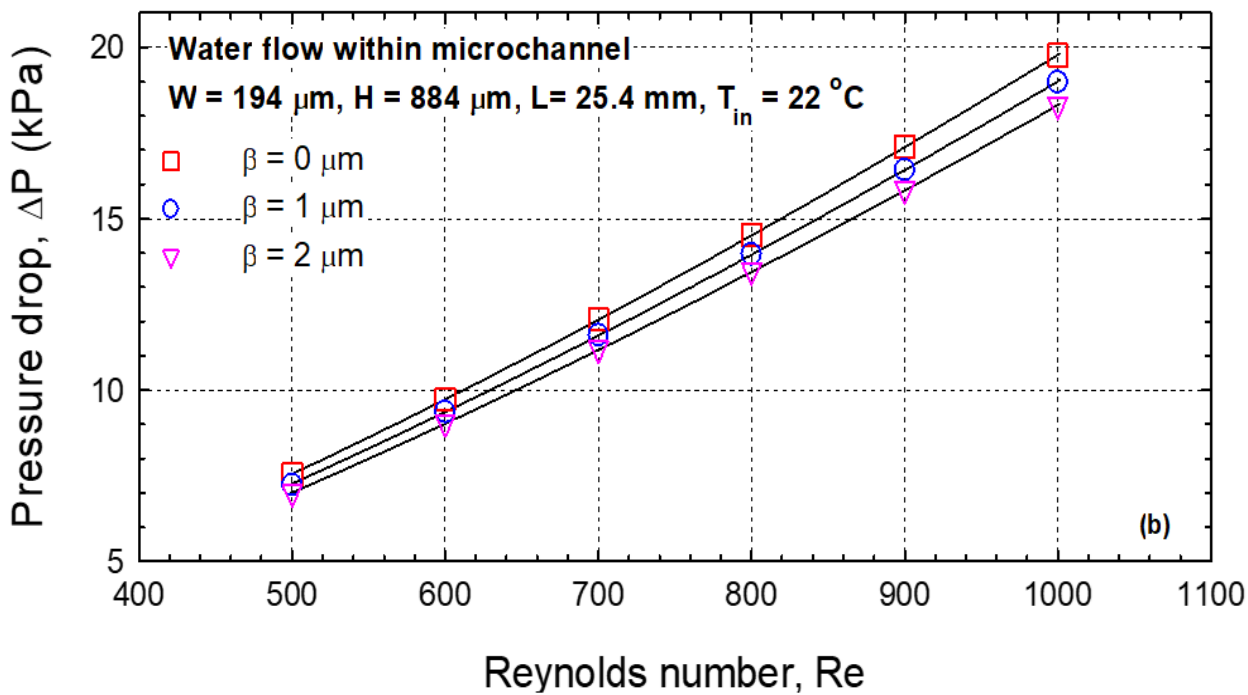

Figure5: Effect of induced slip on (a) heat transfer and (b) pressure drop within a microchannel heat sink.

Figure 6 illustrates the local microchannel wall temperatures, $T_W(z)$ that are calculated using Eq. (3) for different slip lengths at Reynolds number of 800. As it can be seen from the figure, the microchannel wall temperature decreases with increased slip length. At constant applied heat flux, higher Reynolds numbers achieved in microchannels with higher slip lengths as can be seen in figure 5 (a). At constant hydraulic diameter and constant applied heat flux, higher Nusselt numbers result in higher convective heat transfer coefficients. Higher heat transfer rate at constant applied heat flux leads to lower wall temperature as observed in figure 6.

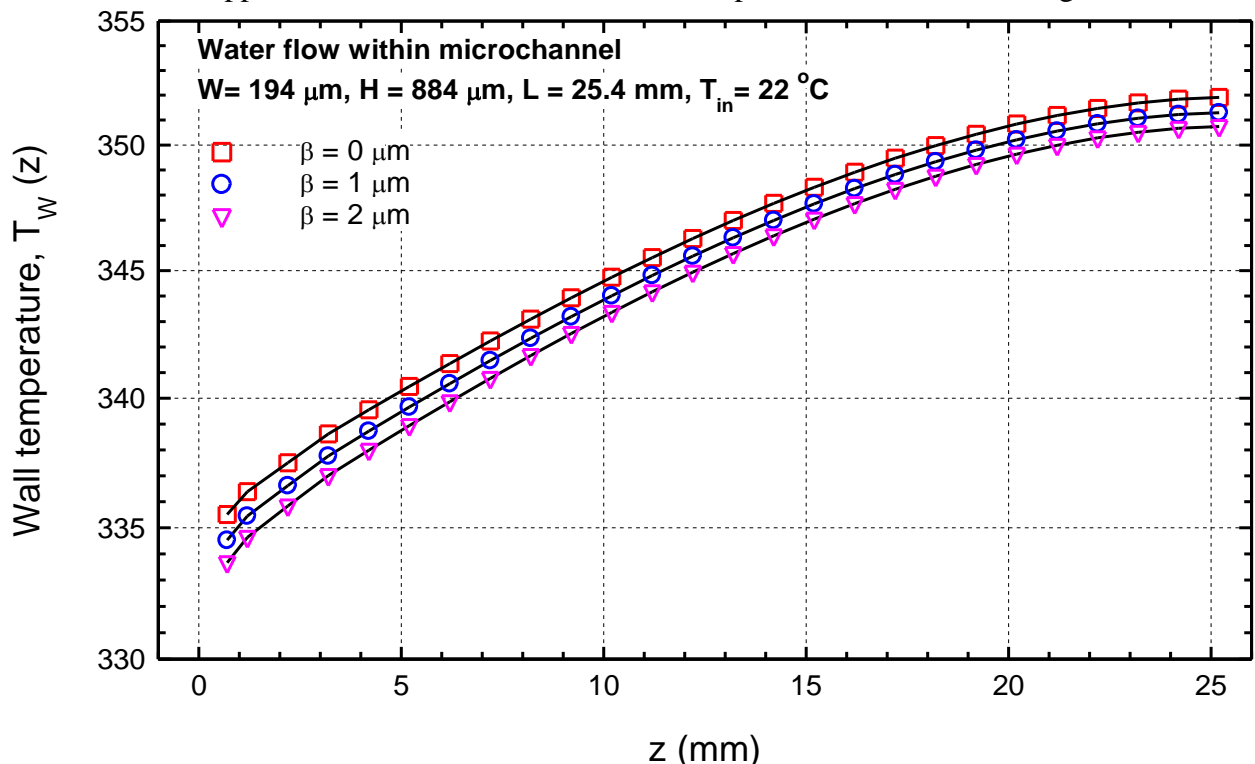

Figure 6: Local wall temperature along the microchannel for different slip lengths at Reynolds number of 800.

## 2.2    Liquid metal flow and heat transfer within miniaturized heat sinks

A miniaturized copper based microchannel heat sink with height of 5 mm, width of 1 mm, and length of 40 mm was considered to evaluate the accuracy of the implemented numerical approach explained in details in section 2 to study the flow and heat transfer of liquid metals within small scale heat sinks. Figure 7 illustrates the finest mesh configuration used in this work based on the V&V 2020 Standard. Total number of 3.8358 million mesh in generated in which 2.43 million mesh elements are in the fluid domain. 15 prismatic layers with growth multiplier of 1.15 are added to the channel walls to accurately capture the velocity and temperature gradients at the interface. Figure 8 shows the maximum wall y⁺ values along the microchannel at Reynolds number of 2300. As it can be seen from the figure 8, the maximum wall y⁺ is below 0.3 for the entire computational domain. Low obtained wall y+ values suggest that the applied mesh grid configuration is even sufficient to estimate velocity and temperature gradient for numerical simulations of turbulent flows. The investigated Reynolds number range in this research is less than 2000 which indicates laminar flow regime. Figure 8 confirms that the chosen thickness of 1.5 µm for the first mesh grid cell adjacent to the wall, keeps the y+ below 1. Also, it indicates that the chosen mesh grid scenario is fine enough to capture the temperature and velocity gradient at the interface.

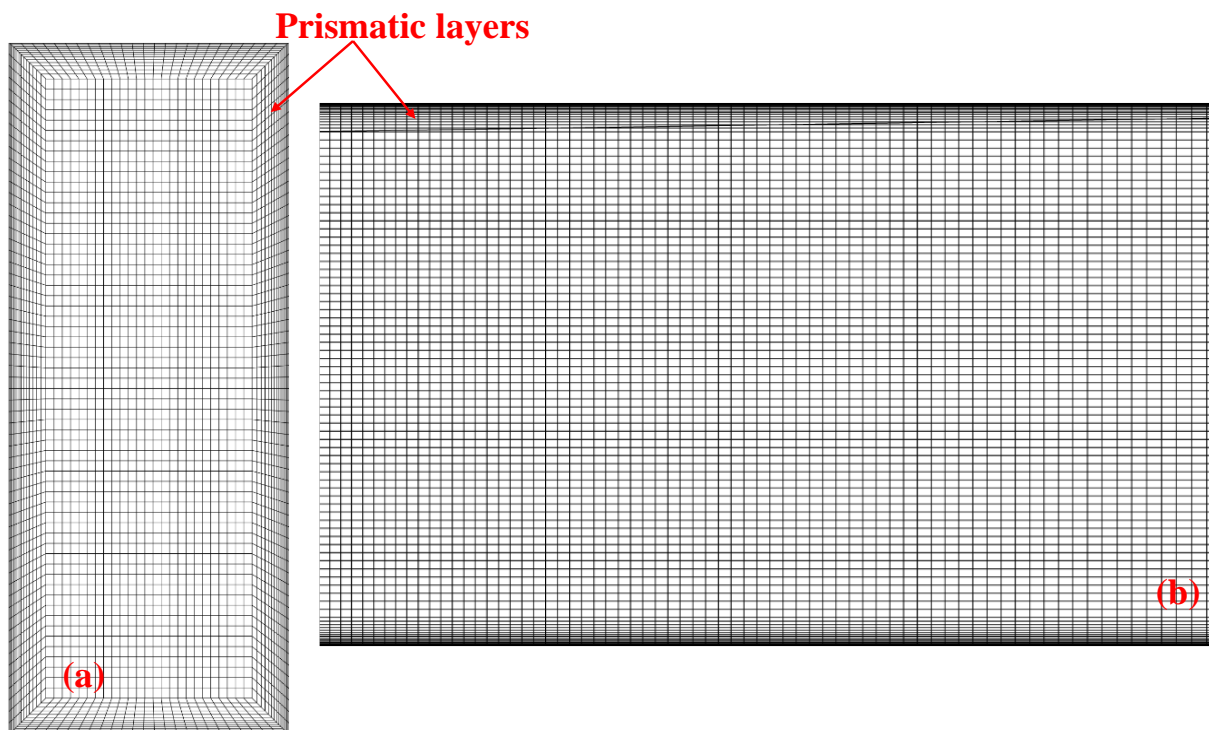

**Prismatic layers**

Figure 7: Mesh grid configuration for (a) channel cross section and (b) in a section along the channel

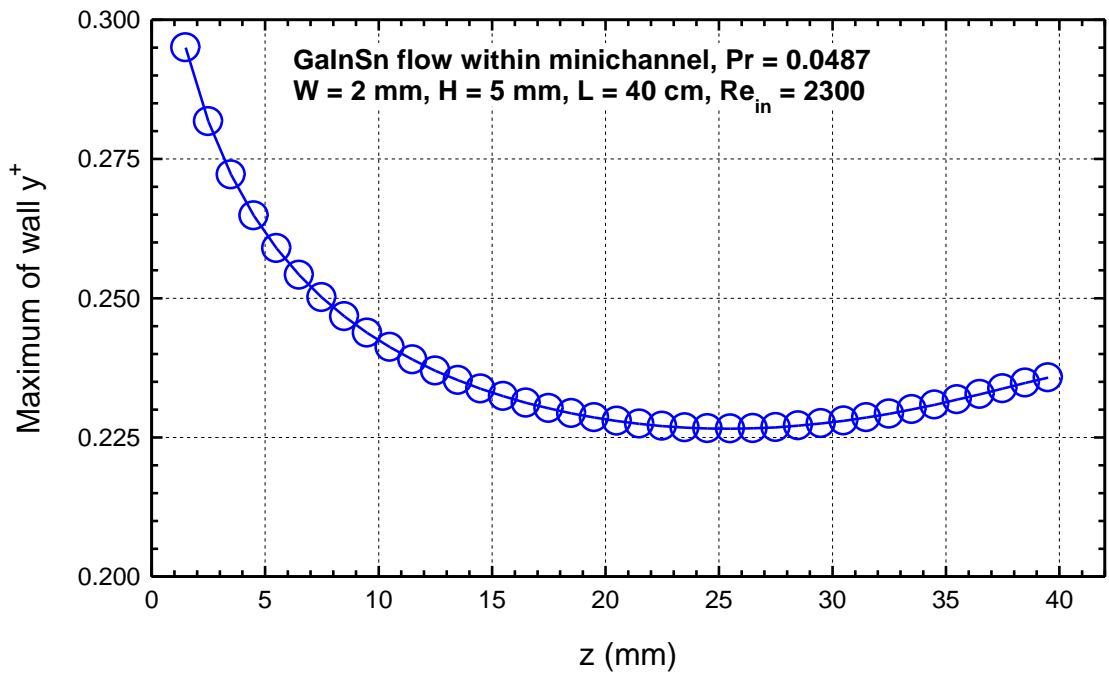

Figure 8: Maximum wall y+ for GaInSn flow a miniaturized heat sink at Reynolds number of 2300

Figure 9 compares the obtained numerical results for the average Nusselt number with experimental data as well as correlations for channel flows [17], [18] and pipe flows [19], [20]. Good agreement is achieved between numerical results and experimental data as shown Figure 9.

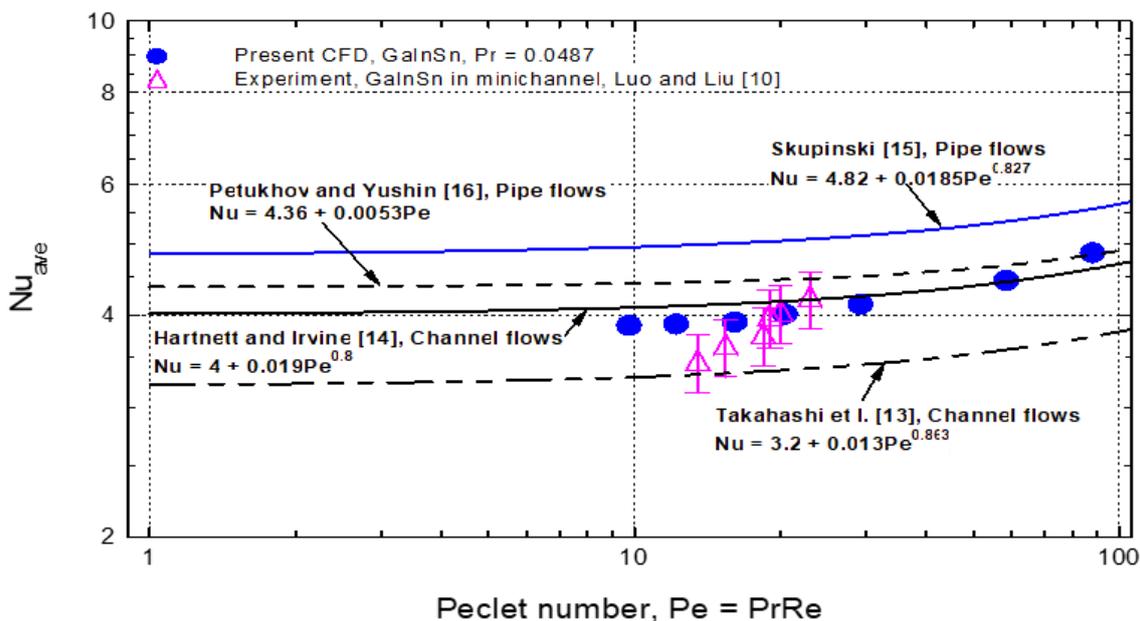

Figure 9: Comparison between numerical results and experimental data for the average Nu number of GaInSn flow in a miniaturized heat sink.

The calculated average Nusselt number increases slowly with increased Peclet number in the laminar flow regime and was within the range of correlations presented in the literature for channel and pipe flows. Figure 8 also verifies the reliability and accuracy of the implemented numerical approach to study forced convection of liquid metals in small scale heat sinks. Another set of numerical simulations were performed to investigate flow and heat transfer of NaK in stainless steel microchannel heat sinks with aspect ratios ($\alpha = H_c/W_c$) of 1 and 2. All the thermophysical properties of the working fluid, NaK, vary with temperature. The first microchannel heat sink had an aspect ratio of 1, width of 400 µm and hydraulic diameter of 400 µm while the second microchannel heat sink had an aspect ratio of 2, width of 200 µm, height of 400 µm and hydraulic diameter of 267 µm. The NaK inlet temperature was set 27 ºC. Figure 10 shows that average Nusselt number for both microchannel heat sinks increase slowly with increased inlet Reynolds number in the laminar flow regime. The Peclet number in Figure 10 is less than 100 and therefore the effect of heat conduction is still noticeable in comparison with the convective heat transfer mechanism. As a result, the Nusselt number becomes a weak function of Reynolds number under laminar flow. Obtained friction factors are also plotted against inlet Reynolds number in Figure 10 (b). The theoretical friction factors for isotropic fully developed laminar flows in channels with aspect ratios of 1 and 2 are also shown in Figure 10 (b).

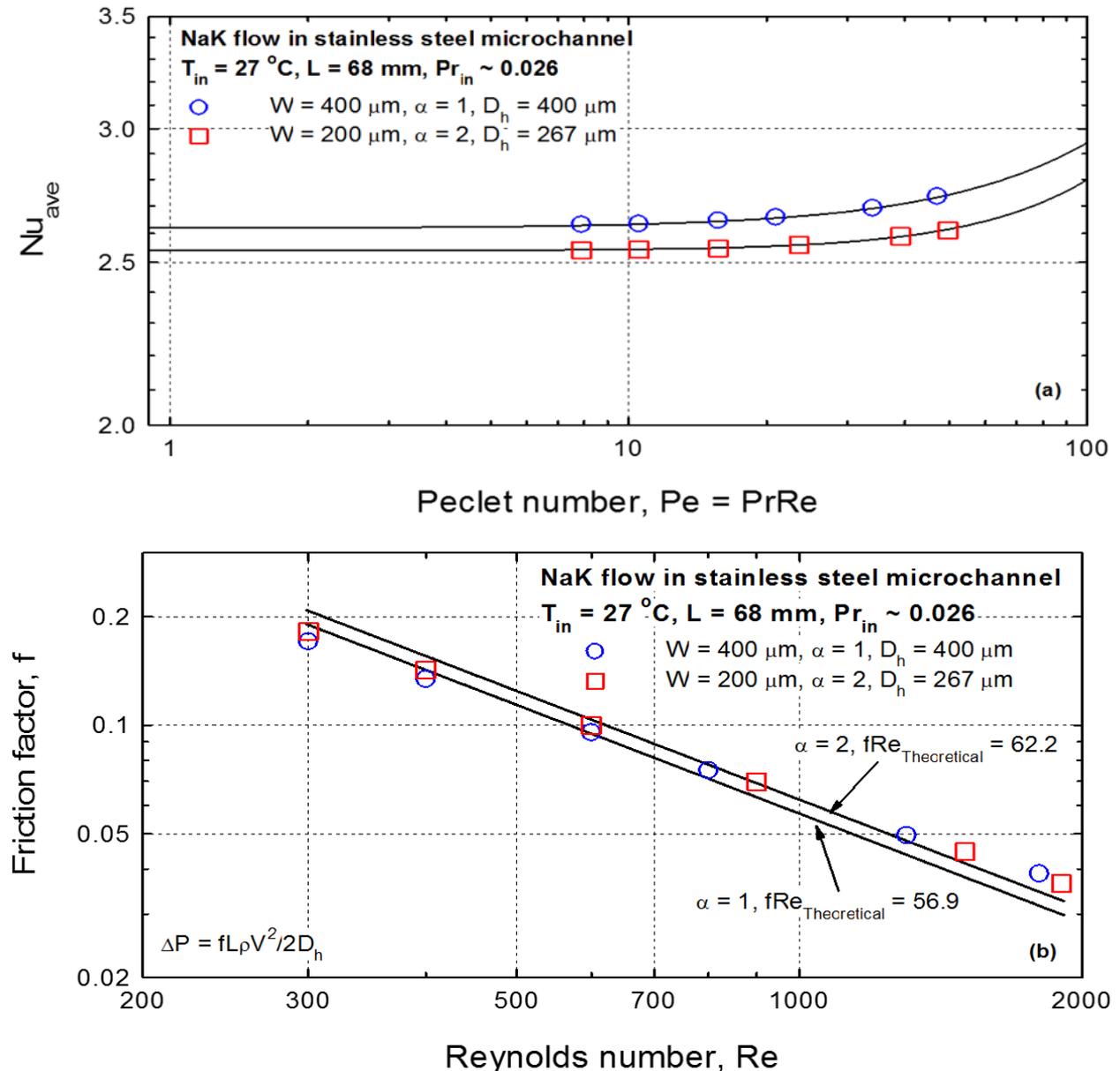

Figure 10: NaK flow/heat transfer within stainless steel microchannel heat sinks. (a) average Nu number (b) friction factor

For low Reynolds numbers, the hydrodynamic entrance region is small while heat transfer at microchannel walls tends to decrease liquid viscosity adjacent to the walls. This reduces total pressure loss along the microchannel heat sink. As a result, the friction factors at Reynold numbers smaller than 500 are lower than the theoretical values. On the other hand, for higher Reynolds numbers, the effect of hydrodynamic entrance region becomes dominant and higher friction factors are obtained in comparison with theoretical values.

## CONCLUSION

Numerical simulations have been performed to study water and liquid metals flows and heat transfer within microchannel heat sinks. The accuracy and reliability of the applied numerical approach has been verified by comparing the obtained numerical results with experimental data and analytical solutions. Besides the no-slip boundary condition, the effect of induced slip at the microchannel walls on heat transfer and pressure drop within the heat sink was also investigated. Results showed that induced apparent slip length of 2 µm can decrease pressure drop by almost 8% while enhancing the convective heat transfer coefficient by 4%. Further numerical analyses were performed to study NaK liquid metal flow and heat transfer within small scale heat exchangers. The Nusselt numbers for forced convection of NaK for the investigated microchannel heat sinks were observed to increase slowly with increased Peclet number. This phenomenon indicated the importance of the heat conduction mechanism in laminar flow of liquid metals within miniaturized heat sinks due to their large thermal conductivity coefficients.

## ACKNOWLEDGMENTS

This paper describes objective technical results and analysis. Any subjective views or opinions that might be expressed in the paper do not necessarily represent the views of the U.S. Department of Energy or the United States Government. Sandia National Laboratories is a multimission laboratory managed and operated by National Technology & Engineering Solutions of Sandia, LLC, a wholly owned subsidiary of Honeywell International Inc., for the U.S. Department of Energy's National Nuclear Security Administration under contract DE-NA0003525.